\documentclass[prl,showpacs,twocolumn,preprintnumbers]{article}
\usepackage{amsmath,bbm,amssymb}
\usepackage{dcolumn}
\usepackage{graphicx}
\usepackage{makeidx}
\usepackage{bm}

\author{Clovis Jacinto de Matos\footnote{ESA-HQ, European Space Agency, 8-10 rue Mario Nikis,
75015 Paris, France, e-mail: Clovis.de.Matos@esa.int}}

\title{Are vortices in rotating superfluids breaking the Weak Equivalence Principle?}

\begin{document}

\maketitle \begin{abstract} Due to the breaking of gauge symmetry
in rotating superfluid Helium, the inertial mass of a vortex
diverges with the vortex size. The vortex inertial mass is thus
much higher than the classical inertial mass of the vortex core.
An equal increase of the vortex gravitational mass is questioned.
The possibility that the vortices in a rotating superfluid could
break the weak equivalence principle in relation with a variable
speed of light in the superfluid vacuum is debated. Experiments to
test this possibility are investigated on the bases that
superfluid Helium vortices would not fall, under the single
influence of a uniform gravitational field, at the same rate as
the rest of the superfluid Helium mass.
\end{abstract}

\maketitle

\section{Introduction}
Vortex dynamics was closely studied in a variety of materials
called quantum liquids and solids \cite{Sonin}, including Bose
Einstein Condensates, Type-II superconductors and in rotating
superfluid $^4 He$. Here we will concentrate on the latter system.
The vortex inertial mass in superfluid $^4 He$, has been
extensively discussed by Duan, Thouless and Popov in
\cite{Duan}\cite{Thouless}\cite{Popov}. In contrast the discussion
of the vortex gravitational mass is particularly poor. To the
knowledge of the author, the validity of the weak equivalence
principle for vortices in rotating superfluids has not been tested
so far. The theoretical discussion of this subject is also scarce
since presently the weak equivalence principle cannot be deduced
on a purely theoretical base.

The Weak Equivalence Principle is one of the main foundations of
the theory of general relativity. It means the constancy of the
ratio between the inertial and the gravitational mass $m_i$ and
$m_g$ respectively of a given physical system.
\begin{equation}
\frac{m_g} {m_i}=\iota=Cte\label{e0}
\end{equation}
This implies, in classical physics, that the possible motions in a
gravitational field are the same for different test particles.
Current experimental tests of the weak equivalence principle
\cite{Bassler} \cite{Smith}, indicate that the gravitational and
inertial masses of any classical physical system should be equal
to each other, $m_g/m_i=\iota=1$, within a relative accuracy of
the E\"{o}tv\"{o}s-factor, $\eta(A,B)$ less than $5 \times
10^{-13}$.
\begin{equation}
\eta(A,B)= (m_g/m_i)_A - (m_g/m_i)_B<5\times 10^{-13}\label{e1}
\end{equation}
However, it should be stressed that until now, all the
experimental tests of the weak equivalence principle have been
carried out with physical systems that do not break gauge
invariance, contrary to the superfluids that break this symmetry.
The E\"{o}tv\"{o}s-factor is usually obtained from the measurement
of the differential acceleration, $\Delta a$, of two free falling
test bodies, $A$ and $B$.
\begin{equation}
\eta(A,B)=\frac{\Delta a}{g_0} \label{e2}
\end{equation}
where $g_0$ is the Earth's gravitational acceleration.

In this paper we argue that \emph{due to gauge symmetry breaking,
the weak equivalence principle is violated by superfluid
vortices}.

The paper starts with the presentation of the current
understanding of superfluid vortex inertial mass. In section 3 a
variable effective Lorentzian vacuum speed of light in superfluids
associated with a breaking of the weak equivalence principle for
vortices is derived from the conservation of energy. In section 4
we discuss the possibility to measure a breaking of the weak
equivalence principle in superfluids through free fall experiments
with rotating superfluid samples. Finally we close with a
discussion of several physical arguments supporting our central
hypothesis.

\section{Vortex inertial mass in rotating superfluids}
A vortex line in rotating superfluid Helium 4 is a topological
singularity, which consists of a normal core region of the size of
the coherence length $\xi$, and an outside region of circulating
supercurrent. The coherence length can be estimated from the
Heisenberg uncertainty principle.
\begin{equation}
\xi\sim\frac{\hbar}{m_{He}c_s}\label{v1}
\end{equation}
where $m_{He}$ is the bare atomic mass in $^4 He$ and $c_s$ is the
speed of sound in the superfluid. Taking $c_s\sim 2\times 10 ^2
m/s$ \cite{Nozieres} we estimate $\xi\sim 1 {\AA}$. In the
theoretical framework of the classical fluid model the only
obvious contribution to the vortex mass is the core
mass\cite{Baym}.
\begin{equation}
m_{core}=L \pi \xi^2 \rho\label{v2}
\end{equation}
where $L$ is the length of the vortex line, and $\rho=Nm$ is the
density with $N$ the bulk number density of $^4 He$ atoms. This
small vortex mass is usually discarded in the equations of motion
of vortex dynamics since it contradicts experimental data.

Duan in \cite{Duan} shown that due to spontaneous gauge symmetry
breaking in superfluids, the condensate compressibility
contributes to a vortex mass which is much larger than the
classical core mass. He calculates that the vortex inertial mass
turns out to diverge logarithmically with the system size.
\begin{equation}
m_{inertial}=m_{core}\ln \Big( \frac{L}{\xi} \Big)\label{v3}
\end{equation}
Where $L$ is the length of the vortex. For a practical superfluid
system $\ln(L/\xi)\sim20-30$.

The number of vortices $N_v$ appearing in a cylindrical sample of
superfluide $^4 He$ rotating with angular velocity $\Omega$ is
deduced from the quantization of the vortex canonical momentum.
\begin{equation}
N_v=\frac{2\pi R^2 \Omega}{\hbar/m}\label{v4}
\end{equation}
where $R$ is the radius of the superfluid sample. The total
increase of the inertial mass of a rotating superfluid sample with
respect to the same non-rotating sample, is obtained from
eq.(\ref{v3}) and eq.(\ref{v4})
\begin{equation}
\Delta M_{inertial}=N_v m_{core}\Big( \ln\Big(\frac{L}{\xi}\Big)
-1\Big)\label{v5}
\end{equation}
Assuming that the weak equivalence principle is still valid in
superfluids this overall increase of inertial mass should appear
together with a similar increase of the gravitational mass of the
superfluid sample.
\begin{equation}
\Delta M_{inertial}=\Delta M_{gravitational}\label{v6}
\end{equation}
Thus we should observe an increase of the weight of the rotating
superfluid sample with respect to the same sample in the
stationary state. Taking a cylindrical sample of radius $R=1 cm$
and $\ln(L/\xi)\sim20-30$, rotating at $\Omega=1 Rad/s$ in
eq.(\ref{v5}), we estimate that the total increase of
gravitational mass is of the order of $\Delta
M_{gravitational}=10^{-14}-10^{-9} Kg$. Thus the experimental
detection of the associated increase of weight of the overall
sample is a challenging task to perform, that has not yet been
overcome by experimentalists. In summary Until the present date
the weak equivalence principle has not been tested for superfluid
vortices. Since the weak equivalence principle cannot be
demonstrated on a purely theoretical basis, it can only be
justified by experiment, and since it has only been experimentally
investigated for the case of physical systems that do not break
gauge invariance, the assumption that the gravitational and the
inertial mass of a rotating superfluid sample are equal,
eq.(\ref{v6}), is an hypothesis that needs to be carefully
investigated at theoretical and experimental level. Specially
because a breaking of gauge symmetry also makes the superfluid
sample a preferred frame in contradiction with the foundations of
relativistic mechanics. Therefore it does not seem too outrages or
exotic to question the hypothesis that \emph{the increase of
inertial mass of the rotating superfluid sample, due to gauge
symmetry breaking, is appearing together with a corresponding
increase of weight of the same sample}, eq.(\ref{v6}), and to
envisage the physical consequences of a possible breaking of the
weak equivalence principle for superfluid vortices.

\section{Variable vacuum speed of light in superfluids
 and breaking of the weak equivalence principle for superfluid vortices}
The breaking of gauge symmetry makes the superfluid sample a
preferred frame, this should be associated with a speed of light
in the superfluid vacuum different from its classical value $c_0$,
appearing in Lorentz transformations. As demonstrated by Duan and
Popov \cite{Duan} \cite{Popov} the vortex inertial mass can be
expressed in function of the vortex static energy $\epsilon_0$
which is also logarithmically divergent as the sample size.
\begin{equation}
m_{inertial}=\frac{\epsilon_0}{c_s^2}\label{v8}
\end{equation}
where $c_s$ is the speed of sound in the superfluid. Making
eq.(\ref{v3}) equal to eq.(\ref{v8}) we conclude that the
effective Lorentzian speed of light $c_{eff}$ for a rotating
superfluid sample is:
\begin{equation}
c_{eff}= c_s \Big(\ln \Big( \frac{L}{\xi}
\Big)\Big)^{1/2}\label{v9}
\end{equation}
For practical values $c_{eff}\sim 4c_s - 5c_s$. We stress that
$c_{eff}$ should be the value of the speed of light to take into
account when carrying out coordinate transformations between a
frame attached to the superfluid sample and a frame outside the
superfluid sample.

Starting from Mach's principle, which asserts that there is a
connection between the local laws of physics and the large scale
properties of the universe, Sciama in \cite{Sciama} introduced the
relation
\begin{equation}
c^2=\frac{2GM}{R}\label{m1}
\end{equation}
where $R$ and $M$ are the radius and the mass of the universe.
Einstein's relationship linking energy and mass then takes the
form
\begin{equation}
E=mc^2=\frac{2GMm}{R}\label{m2}
\end{equation}
which can be interpreted as a statement that the inertial energy
that is present in any physical object is due to the gravitational
potential energy of all the matter in the universe acting on the
object. Therefore the mass $m$ appearing in eq.(\ref{m2}) should
be the gravitational mass of the object.
\begin{equation}
E=m_{gravitational} ~c_0^2\label{m3}
\end{equation}
Since the rest mass energy of the vortex $\epsilon_0$ must be
conserved independently of the value of the vacuum speed of light,
the gravitational mass will adjust its value to compensate the
variation of the speed of light in the superfluid vacuum.
\begin{equation}
m_{gravitational} ~c_0^2=m_{core}c_s^2
\ln\Big(\frac{L}{\xi}\Big)\label{m4}
\end{equation}
From eq.(\ref{m4}) we deduce that the gravitational mass of a
superfluid vortex $m_{gravitational}$ is proportional to the
classical vortex core mass and also diverges logarithmically as
the size of the vortex.
\begin{equation}
m_{gravitational}=\Big(\frac{c_s}{c_0} \Big)^2 m_{core}\ln \Big(
\frac{L}{\xi} \Big ) \label{v7}
\end{equation}
where the proportionality coefficient is equal to the square of
the ratio between the speed of sound in the superfluid $c_s$ and
the classical speed of light in vacuum $c_0$. Comparing
eq.(\ref{v3}) and eq.(\ref{v7}) we conclude that due to the
principle of energy conservation and to the breaking of guage
invariance in superfluids the inertial and the gravitational mass
of a vortex cannot be equal to each other. Therefore the weak
equivalence principle should break for the case of superfluid
vortices.

\section{Rotating superfluids in free fall}
As we have shown in section 2, measuring the vortices
gravitational mass comparing the weight of the superfluid sample
in rotating and stationary state is challenging due to the
extremely small value of the vortex core mass. However in free
fall experiments with rotating superfluid samples it should be
possible to measure the differential acceleration $\Delta a$
between the vortex and the bulk superfluid. The E\"{o}tv\"{o}s
factor $\eta$ associated with the free fall of a vortex and the
superfluid bulk under the single influence of the Earth
gravitational field $g_0$ would be obtained from eq.(\ref{e2}):
\begin{equation}
\eta=\frac{\Delta a}{g_0}
\end{equation}
Let us assume that the friction force between the vortex and the
superfluid bulk is null. On one side, since the superfluid bulk
inertial and gravitational mass are equal, the center of mass of
the superfluid bulk will fall with and acceleration
\begin{equation}
a_{superfluid}=g_0\label{vv1}
\end{equation}
On the other side The vortex will fall according to the equation
of motion
\begin{equation}
g_0 \;  m_{gravitational}=m_{inertial} \; a_{vortex} \label{v9}
\end{equation}
substituting eq.(\ref{v7}), and eq.(\ref{v3}) in eq.(\ref{v9}) we
calculate the vortex falling acceleration
\begin{equation}
a_{vortex}=g_0 \frac{c_s}{c}\label{vv2}
\end{equation}
Substituting the accelerations $a_{superfluid}$, eq(\ref{vv1}),
and $a_{vortex}$, eq.(\ref{vv2}), in eq.(\ref{e2}) we obtain the
E\"{o}tv\"{o}s factor $\eta$ for a superfluid vortex with respect
to the superfluid bulk.
\begin{equation}
\eta=1-\Big(\frac{c_s}{c_0}\Big)^2 \label{v10}
\end{equation}
Taking $c_s\sim2\times 10^2 m/s$ we have $\eta\sim 1$ which is
much higher than the upper limit measured for classical material
systems of $5\times 10^{-13}$, eq.(\ref{e1}). By comparing the
equation of motion of the vortex and superfluid bulk we can easily
show that the relative displacement between the vortex center of
mass and the superfluid bulk center of mass is equal to the
E\"{o}tv\"{o}s factor $\eta$
\begin{equation}
\frac{\Delta d}{d}=\eta\label{v11}
\end{equation}
where $d$ is the distance covered by the superfluid bulk center of
mass after a free fall time $t$. Since the ceiling of the
cylindrical container, enclosing the superfluid, will not allow
the vortex to escape the container we deduce that the vortex
length will shrink while the rotating container is in free fall.
In the most optimistic case the vortex will only disappear when
the container has fall a distance comparable to half its total
length $d=L/2$(the vortex would then be flatten to the ceiling of
the rotating container). At this instant the vortex angular
momentum would be restituted to the rotating container, followed
by the creation of a new vortex with length $L$. Therefore the
rotating container would exhibit a total angular momentum varying
in time with a period $T=\sqrt{L/g_0}$, equal to the period of a
classical pendulum with length $L$, and vortices would be created
an annihilated with the same period of time. In the case where the
weak equivalence principle is preserved superfluid vortices and
the superfluid container angular momentum should not be affected
by the free fall.

If instead of assuming no friction between the vortices and the
superfluid bulk, like we did above, we assume an ideal rigid
connection between both systems. We deduce from the equation of
motion of the freely falling rotating superfluid sample, a falling
acceleration $a_z$.
\begin{equation}
a_z=
\frac{1+\Big(\frac{c_s}{c}\Big)^2\frac{m_v}{m}}{1+\frac{m_v}{m}}g_0\label{v12}
\end{equation}
where $m$ is the total classical mass of the superfluid bulk
(without the vortices) and $m_v=N_v m_{core}
\ln\Big(\frac{L}{\xi}\Big)$ is the total inertial mass of vortices
in the superfluid sample, with $N_v$ being the total number of
vortices. Comparing this acceleration with the falling
acceleration of the same non-rotating sample, $g_0$, we calculate
the E\"{o}tv\"{o}s factor $\eta'$ of the rotating sample with
respect to the non-rotating one.
\begin{equation}
\eta'=\frac{g_0 - a_z }{g_0} \label{v13}
\end{equation}
substituting eq.(\ref{v12}) in eq.(\ref{v13}) we obtain
\begin{equation}
\eta'=\frac{m_v}{\Delta m}\eta\label{v14}
\end{equation}
where $\Delta m=m-m_v$ and $\eta=1-\Big(\frac{c_s}{c_0}\Big)^2$ is
the E\"{o}tv\"{o}s factor of one vortex with respect to the
superfluid bulk (assuming no friction between both),
eq.(\ref{v10}). Taking a cylindrical sample of radius $R=1 cm$ and
$\ln(L/\xi)\sim20-30$, rotating at $\Omega=1 Rad/s$ in
eq.(\ref{v14}), we estimate the order of magnitude of $\eta'\sim
10^{-11}$, which is 2 orders of magnitude above the upper limit
experimentally determined for normal materials, which do not break
gauge invariance, eq.(\ref{e1}). From eq.(\ref{v11}), which also
applies in the present free fall experiment, $\eta'=\Delta d/ d$,
with $\eta'\sim 10^{-11}$, we deduce that large free fall
distances $d$ are required to clearly measure this effect, i. e.
to measure a detectable $\Delta d$ between the distance covered by
the rotating and the non-rotating sample. This could actually be
achieved by conducting the experiment on board a satellite in free
fall around the Earth.

\section{Discussion and Conclusions}
In \cite{Thouless} Thouless and Anglin shown how to obtain an
expression for the inertial mass of a stable quantized vortex in
an infinite neutral superfluid by subjecting it to a straight,
circularly symmetric, pinning potential which is slowly and
steadily rotated about an axis parallel to the vortex line whose
distance from the vortex is large compared with the size of the
vortex. They find that the vortex mass depends strongly on the
pinning potential, and diverges when its radius tends to zero. If
we consider an hypothetical gravitational pinning potential
generated by an adequate distribution of mass, we would thus reach
the conclusion that the vortex inertial mass would depend in a
divergent manner on the gravitational mass of the source of the
pinning potential. Therefore in general the vortex inertial mass
would not cancel out in the equation of motion of the vortex
relative to the pinning potential. This represents an additional
argument in favor of breaking of the weak equivalence principle
for superfluid vortices.

In analogy with the present discussion for superfluids,
spontaneous breaking of gauge invariance in superconductors could
also lead to a breaking of the weak equivalence principle for
Cooper pairs \cite{de Matos}. In this case also a vacuum Lorentz
speed of light different from the classical one, is needed to
preserve energy conservation of the Cooper pairs rest-mass energy.
The anomalous Cooper pair inertial mass excess reported by Tate et
al. \cite{Tate} would thus be the analog in superconductors of the
diverging inertial mass of a vortex in superfluids.

More generally, as widely discussed in the literature \cite{QM}
the validity of the weak equivalence principle in the framework of
quantum mechanics is not at all granted. Since superconductivity
and superfluidity are macroscopic quantum effects, it should not
be too surprising to find some anomalies with the weak equivalence
principle in these systems. Ultimately, it seems to the author,
that this subject should find a general solution in the larger
problem of the correct physical interpretation of quantum and
relativistic mechanics. Meaning that if the Copenhagen
interpretation of quantum mechanics is correct we should indeed
expect a breaking of the weak equivalence principle in physical
quantum systems.

From the analysis presented in section 2 and 4, we see that due to
the small value of the vortex core mass, which is in the order of
$m_{core}\sim10^{-20}-10^{-15} Kg$ (taking $\ln(L/\xi)\sim20-30$),
and due to a speed of sound in the superfluid which is much lower
than the classical speed of light in vacuum, $c_s\ll c_0$, a
possible breaking of the weak equivalence principle for vortices
would not be easy to detect either through weighting measurements
of the rotating and non rotating sample or through the measurement
of anomalous free fall delay times between two superfluid samples
one being rotating, the other being stationary. The most
accessible physical parameters to investigate a possible breaking
of the weak equivalence principle for superfluid vortices would be
the monitoring of the superfluid container angular momentum,
together with the area density of vortices, while the rotating
sample is in free fall. An harmonic oscillation of this parameters
with period $T=\sqrt{L/g_0}$ should be observed.

To conclude, it seems that the current theoretical understanding
of superfluid vortex inertial mass resulting from a breaking of
gauge invariance in superfluids, justifies a careful investigation
of the validity of the weak equivalence principle for superfluid
vortices in the context of free fall experiments with rotating
superfluid samples carried out on Earth or in space.

\end{document}